\documentclass[preprint,pra]{revtex4}
\usepackage{graphicx}
\usepackage{epsfig}
\usepackage{subfigure}
\begin{document}
\title{ Experimental  test of quantum no-hiding theorem}
\author{Jharana Rani Samal 
\footnote{Deceased on her 27th birthday 12th Nov. 2009. The experimental work 
of this paper was completely carried out by the first author. We dedicate this 
paper to the memory of the brilliant soul of Ms. Jharana Rani Samal. }  }

\affiliation{Dept. of Physics and NMR Research Centre,Indian Institute of
Science,Bangalore, India}

\author{Arun K. Pati}
\affiliation{Institute of Physics, Sainik School Post,
Bhubaneswar-751005, Orissa, India}

\author{Anil Kumar}
\affiliation{Dept. of Physics and NMR Research Centre,Indian
Institute of Science,Bangalore,India}

\date{\today}

\begin{abstract}
{\bf
Linearity and unitarity are two fundamental tenets of quantum theory.
Any consequence that follows from these must be respected in
the quantum world. The no-cloning theorem \cite{wz} and the no-deleting
theorem \cite{pb} are the consequences of the linearity and the unitarity.
Together with the stronger no-cloning theorem they provide
permanence to quantum information \cite{rj}, thus, suggesting that 
in the quantum world information can 
neither be created nor be destroyed. In this sense quantum information is 
robust, but at the same time it is also fragile because any interaction 
with the environment may lead to loss of information. Recently, another 
fundamental theorem
was proved, namely, the no-hiding theorem \cite{bp} that addresses precisely 
the issue of information loss. It says that if any
physical process leads to bleaching of quantum information from the
original system, then it must reside in the rest of the universe
with no information being hidden in the correlation between these two
subsystems. This has applications in quantum teleportation \cite{sam}, 
state randomization \cite{sam1}, private quantum channels \cite{amba}, 
thermalization \cite{sandu} and black hole 
evaporation \cite{sh}. Here, we report experimental test of the no-hiding 
theorem with the technique of nuclear magnetic resonance (NMR). We use the
quantum state randomization of a qubit as one example of the bleaching process
and show that the missing information can be fully recovered up to local
unitary transformations in the ancilla qubits. Since NMR offers
a way to test fundamental predictions of quantum theory using coherent
control of quantum mechanical nuclear spin states, our experiment
is a step forward in this direction.}

\end{abstract}

\maketitle

There are many physical processes in nature which lead to apparent
loss of information. The examples can be cited starting from quantum state
randomization to thermalization, black hole evaporation
and so on. For example, in the thermalization process a physical system 
starts from an arbitrary state and after the interaction with a heat bath 
it thermalizes to a canonical thermal distribution. 
Similarly, in the black hole evaporation a pure quantum state
transforms to a thermal mixed state that has no information about the
original pure state \cite{sh}.
Another situation where information about the original system may be lost is
the decoherence \cite{zurek}. 
The process of decoherence is ubiquitous in the quantum world. When a 
quantum system in a pure state interacts with the environment they get 
entangled and in this process the original system looses its purity and 
becomes a mixed state. Typically, the system
looses its phase coherence, but it might happen that after interaction with
the environment the system ends up being in a mixed state that has no 
information about the original. 
If the original information about the system has disappeared
then one may wonder where it has gone? This question has bothered many
scientist. In the classical world, if information is
missing from some system then it may happen in two ways. 
Classical information can be completely hidden from a subsystem if
it is moved to another location, or it may be encoded as
correlations between a pair of subsystems like in the Vernam cipher 
(one-time pad) \cite{vernam}. In general, any information
at classical level will be hidden as a combination of these two \cite{shannon}.
But we know that at a fundamental level quantum theory is at work.
Using the laws of quantum theory it can be shown that if the information is
missing from one system then it simply goes and remains in the rest of
the subsystem which may be regarded as the environment or the universe.
The missing information cannot be hidden in the correlations between the 
system and the environment.
Unlike classical information, quantum theory allows only one way to
completely hide quantum information, that is by moving it to
the remaining subsystems \cite{bp}.

Any physical process that bleaches out the original information we call
it a `hiding' process. Consider a physical process which transforms an
arbitrary pure state
$\rho = |\psi \rangle \langle \psi |$ to a fixed mixed state $\sigma$
that has no dependence on the input state.
Thus, the hiding
process maps $\rho \rightarrow \sigma $ with $\sigma$ fixed for all $\rho$.
Let $\sigma$ has a spectral decomposition as $\sigma = 
\sum_k p_k |k\rangle \langle k |$, where the $p_k$ are the nonzero 
eigenvalues with $\sum_k p_k =1$ and the orthonormal set 
$|k\rangle$ are the eigenvectors.
 Note that this process is a 
generalization of the Landauer erasure map. In the erasure process any 
pure state is transformed to a fixed pure state $|0\rangle$, i.e., 
$|\psi \rangle \langle \psi | \rightarrow |0 \rangle \langle 0 |$
 (the resetting operation) \cite{land}. In the hiding process if the 
final state is such that we 
have all $p_k$'s are zero except one then we have the erasure process.
Therefore, this generalizes the Landauer erasure to any 
hiding process (not merely erasure).

We know that any physical process can be thought of as a unitary
process in an enlarged Hilbert space by attaching an ancilla.
Therefore, the hiding process can be represented
as a unitary transformation from a pure state $|\psi \rangle$ 
to another pure state
$|\Psi \rangle$ such that ${\rm Tr}_{\rm A}(|\Psi \rangle \langle \Psi |) =
\sigma$. The hiding process that
bleaches out the original information can
be expressed in terms of the Schmidt decomposition of the final state as
\begin{equation}
|\psi\rangle
\rightarrow |\Psi\rangle = \sum_{k=1}^K
\sqrt{p_k}\,|k \rangle \otimes|A_k(\psi)\rangle \;, \label{Xform}
\end{equation}
where $p_k$ are the $K$ non-zero eigenvalues of the density matrix
$\sigma$, $\{|k\rangle\}$ are
its eigenvectors and
$\{|A_k\rangle\}$ are the orthonormal states of the ancilla.
Using the linearity and the unitarity of quantum mechanics one can show 
that the final state must be of the following form \cite{bp}
\begin{equation}
|\Psi\rangle =
\sum_k\sqrt{p_k}\, |k\rangle \otimes ( |q_k\rangle
\otimes|\psi\rangle \oplus 0) \;,  \label{thm1}
\end{equation}
where $\{|q_k\rangle\}$ is an orthonormal set of $K$ states and
$\oplus\, 0$ denotes the fact that we substitute any unused dimensions
of the ancilla space by zero vectors. This tells us that under this hiding map
the pure state $|\psi\rangle$ moves to the ancilla.
Since we are free to swap $|\psi\rangle$ with any other state in the ancilla
using purely local unitary operations in the ancilla Hilbert space,
one can say that any
information about $|\psi\rangle$ that is encoded globally is in fact
encoded entirely within the ancilla. Moreover, the quantum information
contained in $|\psi\rangle$ cannot be encoded in the bipartite 
correlations of the system and the ancilla.
This is the no-hiding theorem in quantum theory \cite{bp}.

The simplest example of a hiding process is the quantum state randomization
where an arbitrary pure state in a $d$-dimensional Hilbert space transforms 
to a completely mixed state, i.e.,
$|\psi\rangle \langle \psi | \rightarrow \frac{I}{d}$, where the final density 
matrix is same for all input states.
The exact randomization of an arbitrary pure
state of dimension $d$ can be performed with an ancilla of
dimension at least $d^2$. For a single qubit, the randomization operation 
can be performed in a unitary manner by attaching an ancilla of Hilbert 
space dimension four (by using two qubits). 
State randomization has cryptographic applications, like, in private 
transmission of quantum information using a shared classical key \cite{amba}. 
For example, if Alice encrypts a message by applying some invertible 
quantum operation to some state and converts it to a fixed state, then 
Eve has no chance of knowing the original state. Bob who has access to the key
can recover the message state by applying the invertible quantum operation.

Consider an arbitrary single qubit state  $|\psi\rangle=
\alpha |0 \rangle + \beta |1 \rangle$, where $\alpha$ can be chosen
to be real and $\beta$ a complex number.
According to the state randomization for any input state of a qubit 
the output state will be a random mixture. In this case 
the physical map is a completely positive map as given by
\begin{equation}
|\psi\rangle \langle \psi |
\rightarrow \frac{1}{4}
\sum_{k=0}^3 \sigma_k |\psi\rangle \langle \psi | \sigma_k
= \frac{I}{2},
\end{equation}
where $\sigma_0 =I$ and $\sigma_k, (k=1,2,3)$ are the Pauli matrices.
The above map can be thought of as a unitary map by attaching two qubits as the
ancilla. The unitary transformation is given by
\begin{equation}
|\psi\rangle |A\rangle
\rightarrow |\Psi \rangle =  \frac{1}{2} \sum_{k=0}^3 \sigma_k |\psi\rangle
|A_k \rangle,
\end{equation}
where $|A_k \rangle$ are orthonormal and the initial state of the 
ancilla is an equal superposition of all computational basis states, i.e.,

\begin{equation}
 |A\rangle =  \frac{1}{2} \sum_{k=0}^3|A_k \rangle
  = \frac{1}{2}(|00 \rangle + |01 \rangle + |10 \rangle + |11 \rangle).
\end{equation}
The unitary operator that realizes the above randomization operation 
is a conditional unitary operation (a three qubit gate) which is given by
\begin{equation}
U=\sum_{k=0}^3 \sigma_k \otimes |A_k \rangle \langle A_k|.
\end{equation}
We can see that if we trace out the
ancilla qubits from the final state given in (4) we do get a
completely mixed state. Now, the important question is
where has the missing information gone that has bleached out from the
original qubit. The no-hiding theorem provides answer to this question.

\begin{figure}[]
\centering
\includegraphics[trim = 20mm 220mm 10mm 25mm, clip, scale=0.6]{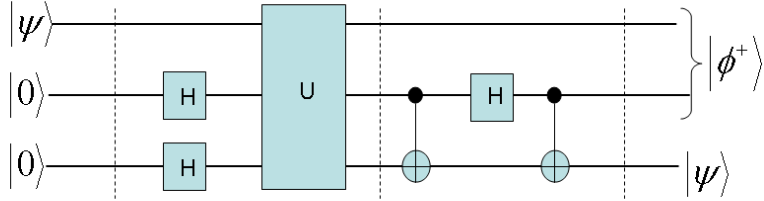}
% Here is how to import EPS art
\caption{Circuit diagram for testing the no-hiding theorem 
using the state randomization.
U is the unitary operator for randomization. H represents the Hadamard gate.
Dots and circles represent the CNOT gates. After randomization, the 
state $|\psi\rangle$ on the first qubit has been transferred to the 
second ancilla qubit (up to local unitary transformation in the ancilla).}
\label{fig1}
\end{figure}

In this simplest example of quantum bleaching process, 
{\em we will see that the missing
information is simply residing in the two qubit ancilla state (up to local
unitary transformation in the ancilla Hilbert space).} This is the essence of
the no-hiding theorem: the hidden
information is fully encoded in the remainder of the Hilbert space.
To reconstruct the original quantum information $|\psi \rangle$ from 
the ancilla qubits we need two CNOT gates
and one Hadamard gate [Fig. 1]. Now, if we apply the following ancilla local
unitary $U_{23}= {\rm CNOT}_{23} (I_2 \otimes  H_{3}) {\rm CNOT}_{23} $ on 
the pure state $|\Psi \rangle$ we get

\begin{eqnarray}
|\Psi \rangle &=&  \frac{1}{2} \sum_{k=0}^3 \sigma_k |\psi\rangle
|A_k \rangle \rightarrow
\frac{1}{2} \sum_{k=0}^3 \sigma_k |\psi\rangle U_{23} |A_k \rangle =
{\rm CNOT}_{23} (I_2 \otimes  H_{3}) {\rm CNOT}_{23}
|\Psi \rangle \nonumber\\
&=& \frac{1}{\sqrt 2} (|00 \rangle + |11 \rangle ) |\psi \rangle =
|\Psi_{\rm out} \rangle.
\end{eqnarray}
From (7) we can see that the first and the second qubits are in the Bell state 
and the third qubit contains the original information.
Thus, the missing information can be fully recovered from the ancilla in 
intact form with no information being hidden in the correlation between 
the system and the ancilla. This shows that the no-hiding theorem is 
indeed respected in the state randomization process.

\begin{figure}[]
\centering
\includegraphics[trim = 45mm 50mm 20mm 25mm, clip, scale=0.7]{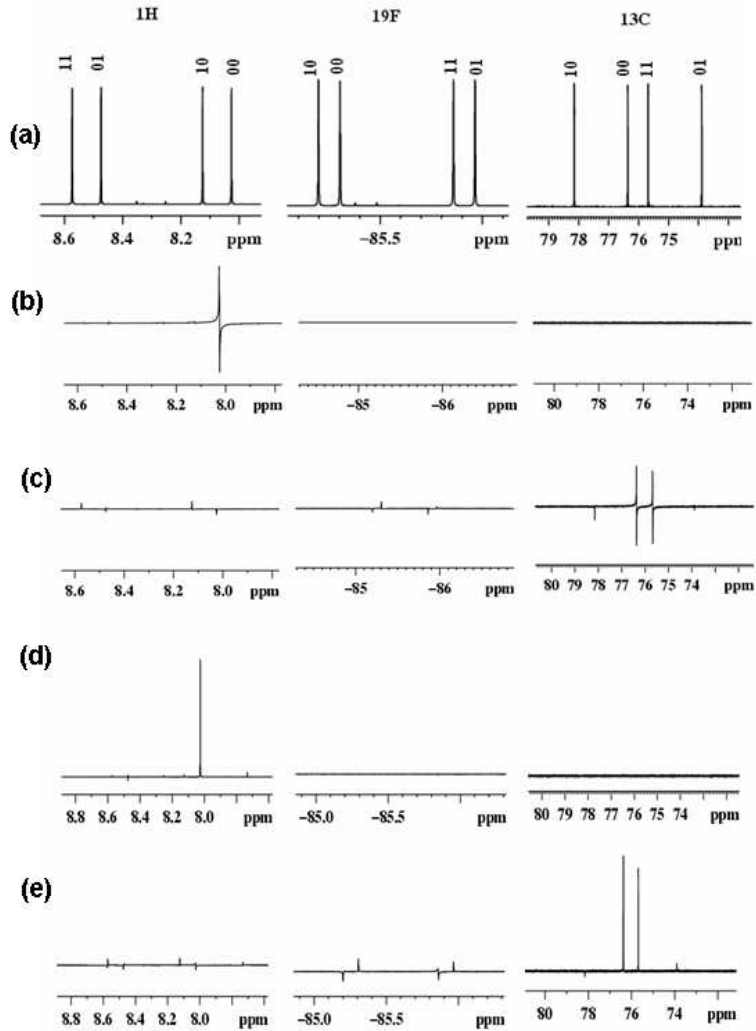}
% Here is how to import EPS art
\caption{(a)The equilibrium spectra of $^{13}CHFBr_{2}$ dissolved
in acetone-d6 at 300K on a AV500 NMR spectrometer. Labelling of the
transition is given on top of each spectral line. The scalar coupling
between different spins are measured as $J_{HC}=224.5$, $J_{FC}=-310.9$,
$J_{HF}=49.7$. For $\theta=\pi/2$  and $\phi=0$, (b) and (c) are respectively 
the experimental spectra for the input
($|\psi\rangle_{1}|00\rangle_{23}$) and output
($|\phi^{+}\rangle_{12}|\psi\rangle_{3}$) state for the 3 spins.
The receiver phase is set using a separate experiment so that y
magnetization appears as positive absorption mode. For $\theta=\phi=\pi/2$,
(d) and (e) are the experimental spectra showing respectively the input and
output states for the 3 spins . }
\label{fig5}
\end{figure}

%{\bf Experimental Implementation:}
In the sequel we give details of the experimental implementation of the 
quantum state randomization.
Liquid state NMR has been successfully used as a test bed for a large number 
of quantum information protocols including Grover's algorithm \cite{grover}, 
Shor's algorithm \cite{shor}, quantum teleportation \cite{tele}, 
adiabatic quantum computation \cite{amitra,peng}, estimation of the 
ground state of Hydrogen atom up to $45$ bits \cite{du} and more recently 
experimental verification of the non-destructive discrimination of 
Bell-states \cite{samal}.
Here, we report an experimental verification of the quantum no-hiding 
theorem using NMR. Experiments have been performed in a three qubit 
hetero-nuclear spin system
formed by the $^{1}H$, $^{19}F$ and $^{13}C$ nuclei of $^{13}C$-enriched 
dibromo fluoro methane ($^{13}CHFBr_{2}$) \cite{mitra}. 
Fig. 2(a) shows the equilibrium spectrum for the three
nuclei at 300K recorded in a Bruker AV500 spectrometer, where the resonance
frequencies of $^{1}H$, $^{19}F$ and $^{13}C$ are 500 MHz, 470 MHz and
125 MHz respectively. We have taken $^{1}H$, $^{19}F$ and $^{13}C$ as the first,
the second and the third qubit, respectively.

\begin{figure}[]
\begin{center}
\subfigure[]{\label{fig:edge-a}\includegraphics[trim = 40mm 30mm 30mm 25mm,
clip, angle=90, scale=0.65]{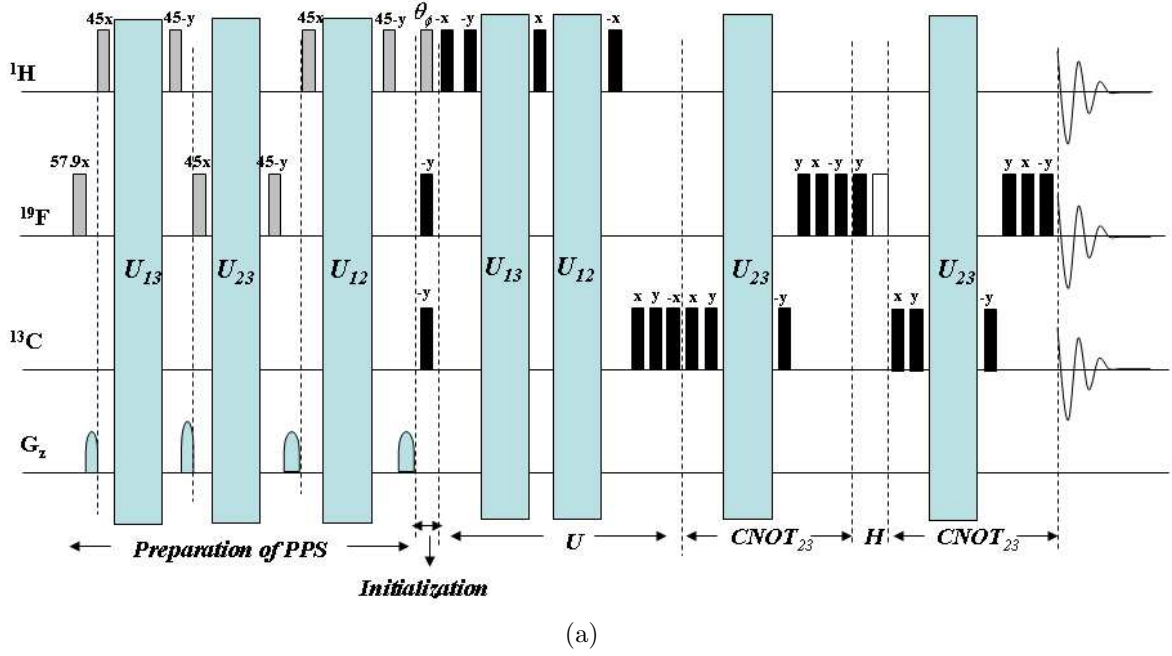}}
\subfigure[]{\label{fig:edge-b}\includegraphics[trim = 85mm 60mm 25mm 30mm,
 clip, angle=90, scale=0.7]{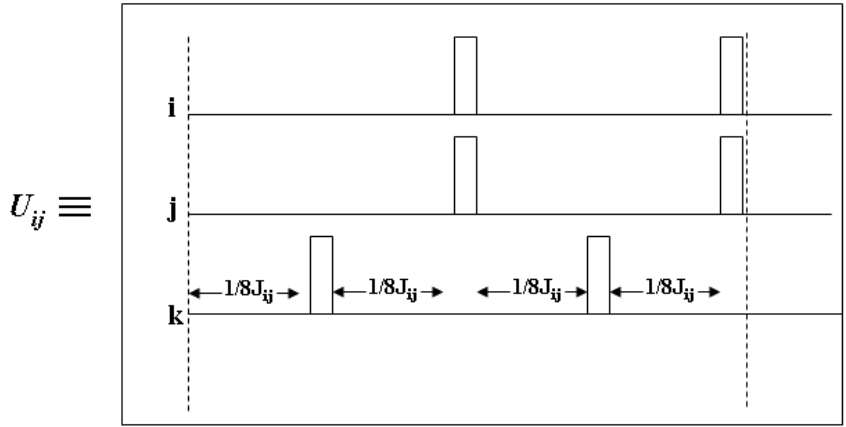}} \\
\end{center}
\caption{(a) NMR pulse sequence for the implementation of the no-hiding theorem
using the state randomization. The filled and the empty boxes respectively
 represent
$\pi/2$ and $\pi$ pulses while the grey boxes represent
pulses with flip angles on the top. The phase of a pulse is given on top
of the pulse for $\pi/2$ and $\pi$ pulses and are subscripts of angles for
other flip angle pulses. Gz is the z-gradient pulse used to destroy all
unwanted transverse magnetization. For each $(\theta,\phi)$ pair three
identical experiments are performed to observe each qubit independently.
(b) The pulse sequence corresponding to the $U_{i,j}$ operator.}
\label{fig:edge}
\end{figure}

Fig. 1 depicts the circuit diagram that implements the state randomization for 
a single qubit in a unitary manner.
Using the quantum circuit of  Fig. 1, an equivalent NMR pulse sequence
has been developed here (Fig. 3).
This pulse sequence contains several elements namely:
(i) the preparation of the pseudo pure state (PPS) \cite{mitra}, 
(ii) the process of initialization, (iii) the randomization operation using its 
unitary extension, 
(iv) extraction of the original quantum information from the ancilla by 
applying local unitary transformations and
(v) finally the measurement (reconstruction of density matrices via tomography).
These sequence of steps can be represented by schematic use of several
$U_{ij}$ blocks (shown in light grey).
During the application of $U_{ij}$ the system evolves under the scalar 
coupling $J_{ij}$ between the spins $i$ and $j$ for a
time period of $1/2J_{ij}$. They create two spin order modes from a
single spin transverse mode and vice versa.
The $U_{ij}$ is shown in an expanded manner in Fig. 3(b). 
The $\pi$ pulses in the center of $U_{ij}$ (Fig. 3(b))
 are used to refocus all the chemical shifts and all the scalar couplings
except between the spins $i$ and $j$.

The experiment consists of implementing the above five steps
on the three qubit system. 
%The preparation of PPS:
The initial part of Fig. 3(a) contains preparation
of  $|000\rangle$ pseudo pure state (PPS) by spatial averaging method 
\cite{cfh}. 
%Next is the initialization: 
The initial state $|\psi\rangle_{1}|A\rangle_{23}$ (see (4))
is prepared from $|000\rangle$ PPS by applying a $(\theta)_{\phi}$  pulse
on the first spin and $[\pi/2]_{-y}$ pulses (Hadamard gates) on the 
second and the third spins, respectively. The $(\theta)_{\phi}$ pulse in NMR is 
represented by the operator

\begin{equation}
 U_{\theta,\phi}=\left(
\begin{array}{cc}
\cos(\theta/2) & -ie^{-i\phi}\sin(\theta/2) \\
-ie^{i\phi}\sin(\theta/2) &\cos(\theta/2)   \\
\end{array}
\right)  
\end{equation}
which transforms the state $|0\rangle$ to the superposition state 
$|\psi\rangle$, i.e.,
\begin{equation}
|0\rangle\rightarrow \cos(\theta/2)|0\rangle+e^{i(\phi-\pi/2)} 
\sin(\theta/2)|1\rangle=\alpha|0\rangle+\beta|1\rangle=|\psi\rangle.
\end{equation}
The Hadamard gate $[\pi/2]_{-y}$ on the second and third qubits transforms
$|00\rangle $ to $|A \rangle $.
Then, we need to perform the randomization operation $U$ which is given by
 \begin{equation}
  U=  \left(
      \begin{array}{cccccccc}
       1 & 0 & 0 & 0 & 0 & 0 & 0 & 0 \\
      0 & 0 & 0 & 0 & 0 & 1 & 0 & 0 \\
      0 & 0 & 0 & 0 & 0 & 0 & 1 & 0 \\
      0 & 0 & 0 & 1 & 0 & 0 & 0 & 0 \\
      0 & 0 & 0 & 0 & 1 & 0 & 0 & 0 \\
      0 & 1 & 0 & 0 & 0 & 0 & 0 & 0 \\
      0 & 0 & -1 & 0 & 0 & 0 & 0 & 0 \\
      0 & 0 & 0 & 0 & 0 & 0 & 0 & -1 \\
      \end{array}
    \right) .
 \end{equation}
Conversion of this matrix into NMR pulse sequence has been achieved here by
the use of a novel algorithmic technique for efficient decomposition of the 
unitary
operators in NMR developed in our laboratory by Ajoy et al \cite{ajpa}. This
method uses  graphs of a complete set of base operators and develops an
algorithmic technique for finding the decomposition of a given unitary
operator into basis operators and their equivalent pulse sequences.
The equivalent pulse sequence for the unitary operator given in Eq(10) can be
expressed as
 \begin{equation}
 U=exp(-i\frac{\pi}{4}\textbf{1}) exp(i\frac{\pi}{2}I_{3z}) exp(-i\pi
I_{1y}I_{2z})exp(-i\pi I_{1z}I_{3z})exp(i\frac{\pi}{2}I_{1x})
exp(i\frac{\pi}{2}I_{1z}).
\end{equation}
The first term yields only an overall phase and can be neglected. The
second and last
terms are $\pi/2$ rotations of the third and the first qubits, respectively
about the z-axis. The z-rotation is achieved by the composite pulse
$[\pi/2]_{-z}^{i}=[\pi/2]_{x}^{i}[\pi/2]_{-y}^{i}[\pi/2]_{-x}^{i}$, where
$i (i=1, 2, 3)$ refers to various qubits. The fourth term is
$U_{13}$ and the third term can be converted to $U_{12}$
by converting $I_{1y}$ to $I_{1z}$ using the pulse 
$[\pi/2]_{-x}[\pi/2]_{y}[\pi/2]_{x}$.
The final pulse sequence is obtained as

\begin{equation}
[\pi/2]_{-z}^{3}[\pi/2]_{-x}^{1}U_{12}[\pi/2]_{x}^{1}U_{13}[\pi/2]_{-x}^{1}
[\pi/2]_{-z}^{1}.
\end{equation}
Here, pulses are always applied from right to left.

The next step is extraction of quantum information $|\psi\rangle$ 
from the ancilla qubits. As we have seen 
this can be achieved by applying two CNOT gates and one Hadamard gate 
(Fig. 1 and 3). ${\rm CNOT}_{23}$ gate in NMR is realized by the 
following pulse sequence \cite{ccl}

\begin{equation}
[\pi/2]^{2}_{-y}[\pi/2]^{2}_{x}[\pi/2]^{2}_{y}[\pi/2]^{3}_{-y}U_{23}[\pi/2]^{3}_{y}
[\pi/2]^{3}_{x}.
\end{equation}
Finally, we need to do measurements to confirm our result.
All the qubits are directly observed at the end of the
computation and no measuring pulse is required. The missing information about
$|\psi\rangle$ is actually found in the state of the third qubit. This requires
taking the trace over the first and the second qubits. One of the ways of taking
this trace is to decouple the first and the second qubits while observing
the third qubit \cite{scs}. However, this leads to excessive sample
heating \cite{scs}. We have therefore performed the trace
numerically by measuring all the three qubits and appropriately adding the
signal intensities \cite{cjf}.

The full density matrix of the output state $|\Psi_{\rm out} \rangle$ (with the
states ordered as 000, 001, 010, 011, 100, 101, 110, 111) is given by
\begin{equation}
|\Psi_{\rm out}\rangle \langle \Psi_{\rm out} | = \left(
                                       \begin{array}{cccccccc}
                                         \alpha^{2} & \alpha\beta^{*}& 0 & 0
& 0 & 0 & \
\alpha^{2}  &\alpha\beta^{*} \\
                                         \beta\alpha^{*} & \beta^{2} & 0 & 0
& 0 & 0 &
\beta\alpha^{*} & \beta^{2} \\
                                         0 & 0 & 0 & 0 & 0 & 0 & 0 & 0 \\
                                         0 & 0 & 0 & 0 & 0 & 0 & 0 & 0 \\
                                         0 & 0 & 0 & 0 & 0 & 0 & 0 & 0 \\
                                         0 & 0 & 0 & 0 & 0 & 0 & 0 & 0 \\
                                         \alpha^{2}  & \alpha\beta^{*} & 0 & 0
& 0 & 0 & \alpha^{2}  & \alpha\beta^{*} \\
                                          \beta\alpha^{*}  & \beta^{2} & 0 & 0
& 0 & 0 & \beta\alpha^{*}  & \beta^{2} \\
                                       \end{array}
                                     \right).
\end{equation}
% where $\alpha$, $\beta$ are defined in Eq. 9.
Eq (14) contains two single quantum terms
of amplitude $\alpha\beta^{*}$ on the third qubit ($\sigma_{12}, \sigma_{78}$,
and  complex conjugate (cc)) which are directly observable. Since both the
single quantum coherence of the third spin are represented by
$\alpha \beta^{*}$, they are in phase with each other.
We note that no single quantum (SQ) coherence of spins $1$ and $2$ are 
present as they are in the Bell state.
There are two double quantum (DQ) terms of amplitude $\alpha^{2} \& \beta^{2}$
($\sigma_{17} \& \sigma_{28}$ and cc) and one triple quantum (TQ) term  
of amplitude
$\alpha\beta^{*}$ ($\sigma_{18}$ and cc) which are not directly observable.
They have been observed ( for tomography) by converting them to observable
single quantum term \cite{a}.

\begin{figure}[]
\centering
\includegraphics[trim = 5mm 140mm 8mm 5mm, clip, scale=0.7]{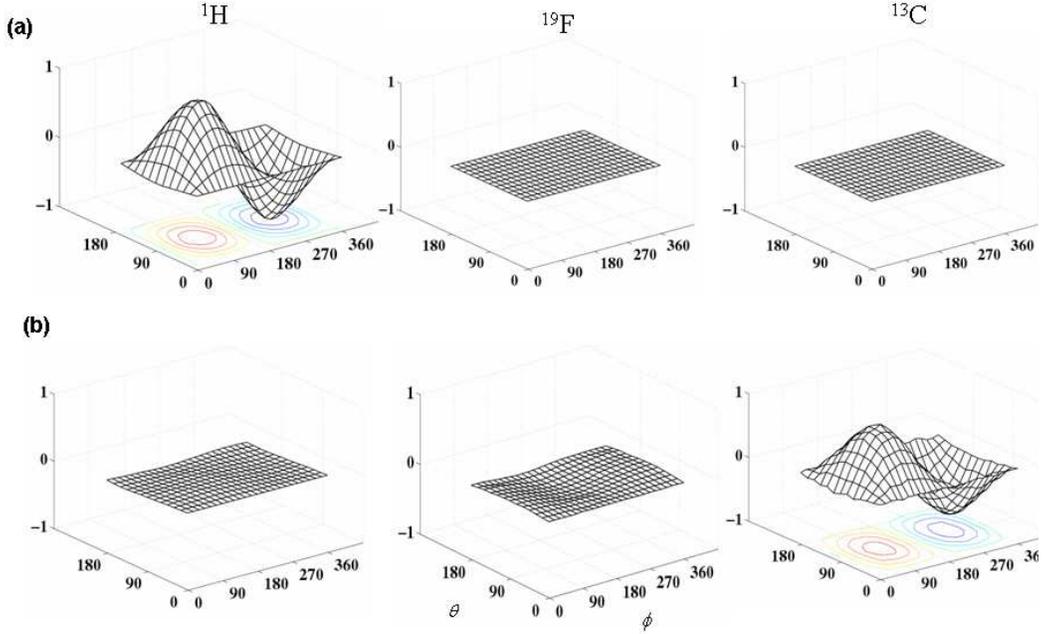}
% Here is how to import EPS art
\caption{Summarized experimental results for the no-hiding theorem using the 
state randomization. The integrals of the real part of the NMR signal from spins
$^{1}H$,$^{19}F$ and $^{13}C$ are shown as mesh and contour plots as a function
of $\theta$ and $\phi$. The plots show the expected sine and cosine behavior.
(a) Input state $|\psi\rangle_{1}|00\rangle_{23}$. The information about
$\theta$ and $\phi$ is encoded in the first spin, and second and third spins 
are in $|0 0 \rangle$ state.
(b) Final output state of three spins 
$1/\sqrt{2}(|00 \rangle_{12} + |11 \rangle_{12}) |\psi\rangle_{3}]$. The
information about the state $|\psi\rangle$ has been transferred 
from the first qubit to the third qubit. The first and second qubits are 
in the Bell state.}
\label{fig6}
\end{figure}

%{\bf Results:}
\begin{figure}[]
\centering
\includegraphics[trim = 10mm 120mm 10mm 10mm, clip, scale=0.6]{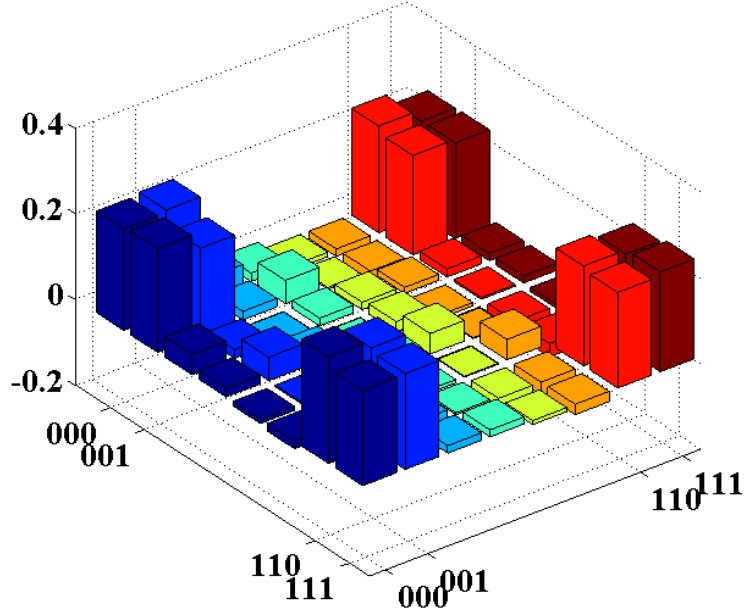}
% Here is how to import EPS art
\caption{Tomographed density matrix of the output state for 
$\theta=\phi=\pi/2$.}
\label{fig1}
\end{figure}

\begin{figure}[]
  \begin{center}
    \subfigure[]{\label{fig:edge-a}\includegraphics[trim = 0mm 125mm 0mm
0mm, clip,
scale=0.35]{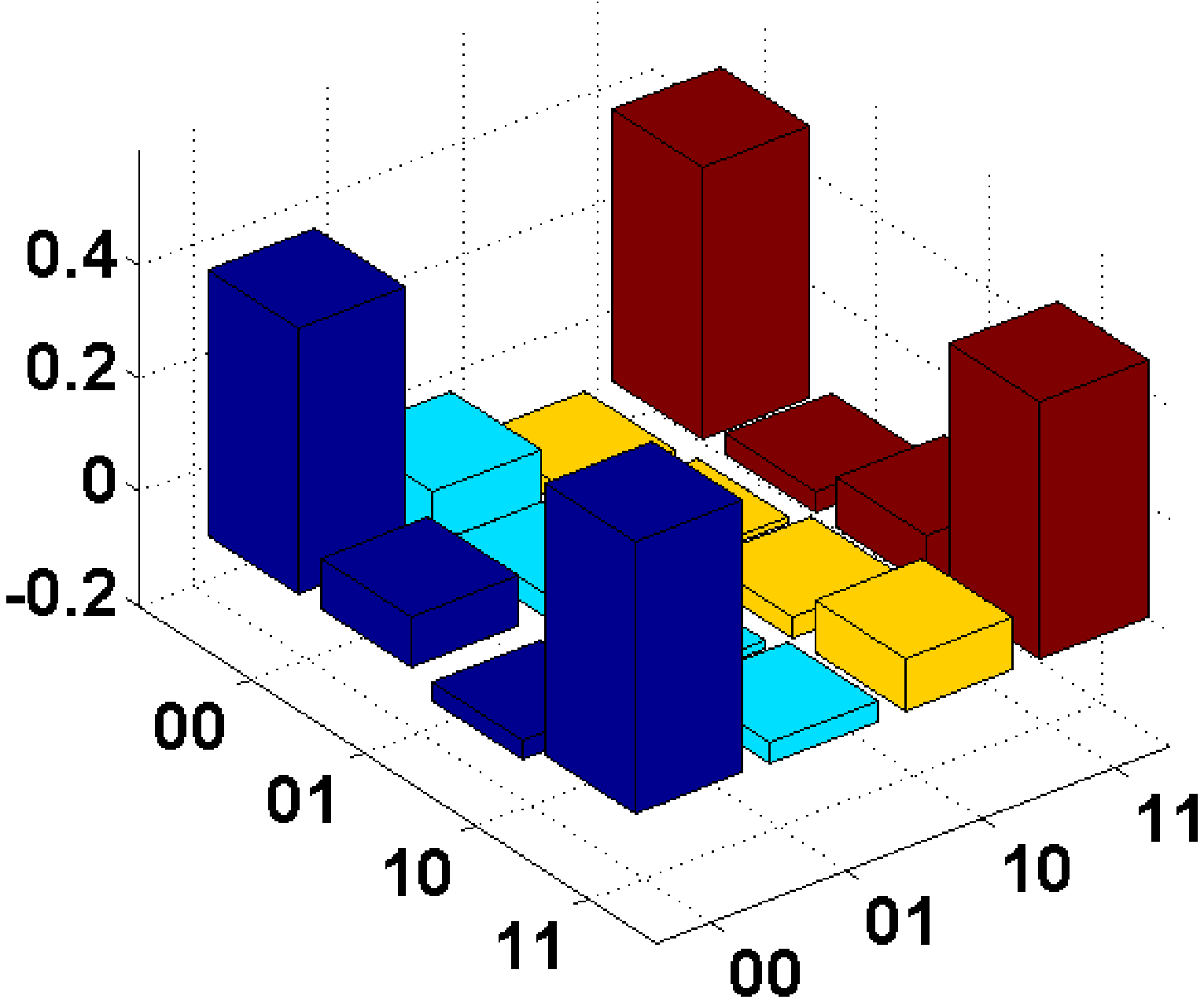}}
 	 \subfigure[]{\label{fig:edge-b}\includegraphics[trim = 0mm 125mm
0mm 0mm,
clip, scale=0.35]{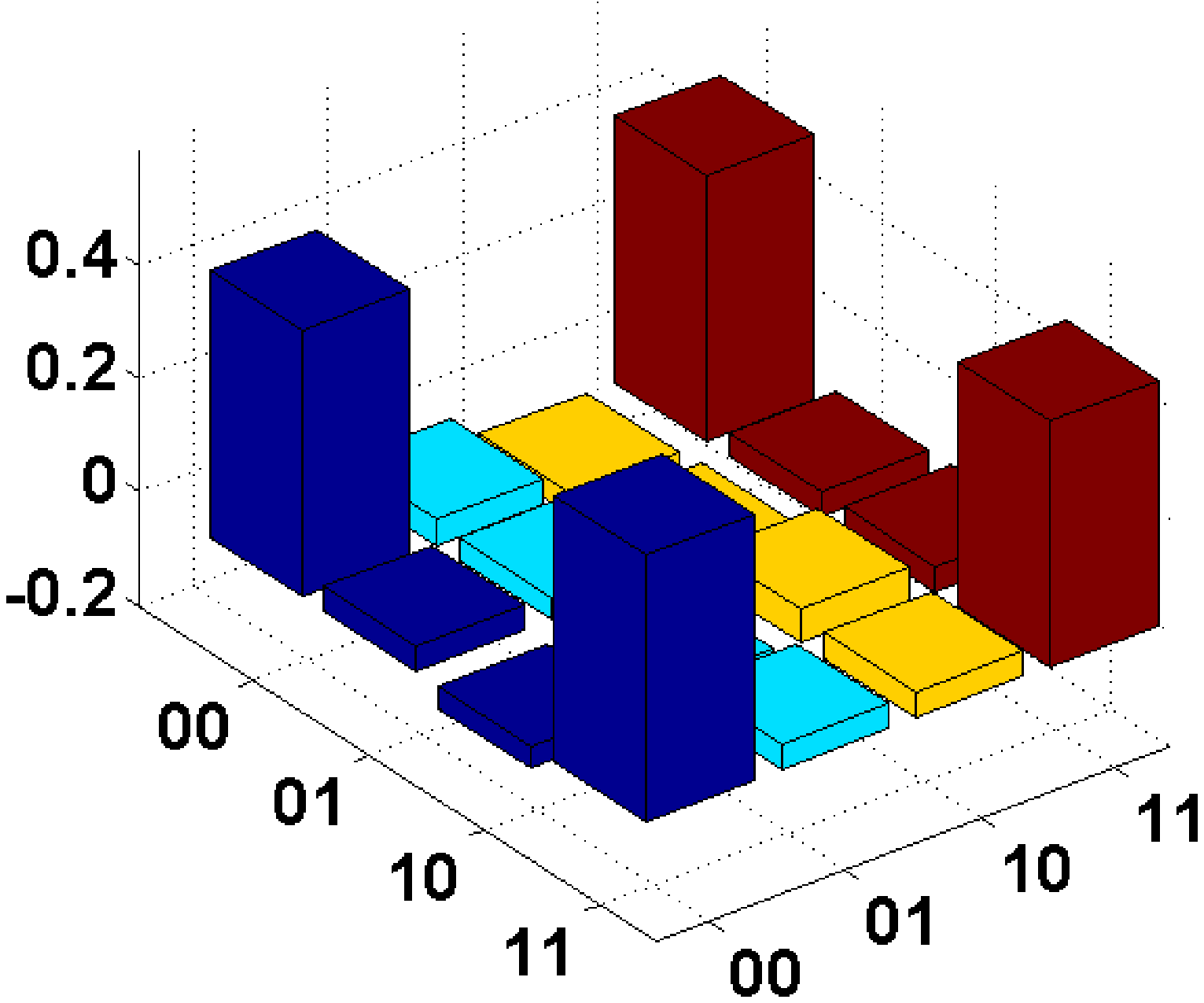}} \\
   \end{center}
  \caption{The output density matrix for the first two qubits showing that they 
are in the Bell state $|\phi^{+}\rangle$. (a) For $\theta = \phi = \pi/2$
  (b) For $\theta = \pi/2$ and $\phi = \pi$}
  \label{fig:edge}
\end{figure}

The results are summarized in Figs (2) and (4). We have measured a
total of 325 input states arranged on a 13 by 25
rectangular grid of $\theta(0 \rightarrow \pi) $ and 
$\phi (0 \rightarrow 2\pi)$ values with a
spacing of $15^{0}$. The experimental
spectra of the input and output states for $\theta = \pi/2$, $\phi = 0$ 
are respectively given in Fig. 2(b) and
Fig. 2(c). Similarly, for $\theta = \phi = \pi/2$ these are 
respectively given in Fig. 2(d) and Fig. 2(e). The receiver phase
is set such that we get positive absorption
lines for the input state $\theta = \phi = \pi/2$. For each input state 
we have separately
measured the total NMR signal (integration of the entire multiplet)
observed from all the spins 1, 2 and 3 and their real components are plotted
in Fig. 4. Fig. 4(a) shows the input state
$|\psi\rangle_{1}|00\rangle_{23}$
 and Fig. 4(b) shows the output state $|\phi^{+}\rangle_{12}|\psi\rangle_{3}$ .
{\it The experimental results (Fig. 4) clearly show sine and cosine modulation
of the expected lines, thus showing the coherent quantum oscillations of the 
original qubit. This 
ensures that the information about $\alpha$ and $\beta$ has been actually 
transferred from the first qubit to the third qubit in accordance with the 
no-hiding theorem.}

   Additionally we have reconstructed the density matrices of the output
state for several values of $\theta$ and $\phi$ by
quantum state tomography \cite{a}. The density matrix of the output state for
$\theta=\phi=\pi/2$ is shown in Fig. 5. The experimental state tomography 
confirms the theoretical output state as given in (14). The fidelity of 
the measurement has been evaluated for several values of
$\theta~ \& ~\phi$ using the parameters ``\emph{average absolute deviation}
$\langle\Delta X\rangle$'' and the ``\emph{maximum absolute deviation}
$\Delta X_{max}$'' as defined by

\begin{equation}
\langle\Delta X\rangle=\frac{1}{N^{2}}\sum_{i,j=1}^N|x_{i,j}^{T}-x_{i,j}^E|,
\end {equation}

\begin{equation}
{\rm and}~~ \Delta X_{max}=Max|x_{i,j}^{T}-x_{i,j}^E|, \forall i,j\in \{1,N\},
\end{equation}
where $x_{i,j}^{T}$, $x_{i,j}^{E}$ are the theoretical and the 
experimental elements \cite{mitra}. The average absolute deviation 
( for three $\theta ~ \& ~ \phi$ values)
$\langle\Delta X\rangle$ was found to be $\sim 2\%$ and the maximum absolute
deviation $\Delta X_{max}$ was found to be $\sim 5\%$.

In our experiment the reduced density matrices of the first two qubits 
of the output states ($|\phi^{+}\rangle$ of Fig. 1) have also been 
tomographed to observe the
fidelity of the Bell states in the first two qubits at the end of the
measurement. Fig. 6(a) contains the Bell state $|\phi^{+}\rangle_{12}$
for the input state with parameters $\theta = \phi = \pi/2$ and 
Fig. 6(b) contains $|\phi^{+}\rangle_{12}$
for the input state with parameters $\theta = \pi/2$, $\phi = 0$. 
Fig. 6 confirms that the first two qubits remain in the Bell state 
irrespective of the changes in
$(\theta, \phi)$ and the original information about $(\theta, \phi)$ 
have been transferred to the third qubit. The average absolute 
deviation $\langle \Delta X \rangle$
was found to be $\sim 5\%$ and
the maximum absolute deviation $\Delta X_{max}$ was found to be $\sim 7\%$.
The experimental errors can originate from rf inhomogeneities, 
imperfect calibration of rf pulses and decoherence. 
However, in the present experiment the decoherence errors 
are likely to be small, since the total experimental time
(Fig. 3) is $\sim 30$ msec while the shortest $T_{2}$ (of $^{19}F$)
of the sample is $700$ msec.

%{\bf Conclusions:}

%In this work we have reported the experimental test of the 
%no-hiding theorem on a $3$-qubit NMR quantum information processor.
To conclude, we have performed a proof-of-principle demonstration of 
the no-hiding theorem  and 
addressed the question of missing information 
on a $3$-qubit NMR quantum information processor.
Using the state randomization as a prime example of the bleaching process we 
have found that the original quantum information which is missing from 
the first qubit, indeed can be fully recovered from the
ancilla qubits. We have encoded the quantum information in 
$^{1}H$ spin and later it has been fully  
reconstructed experimentally from $^{13}C$ spin using the 
NMR pulse sequence. No information is found to be hidden in the bipartite 
correlations between the original qubit and the ancilla qubits. 
To the best of our knowledge this is the first experimental
verification of this fundamental theorem of
quantum mechanics. 

Randomization of quantum state is an important primitive in private quantum 
channels which permits private transmission of quantum information using a 
shared classical key. One can randomize $n$-qubits perfectly using $2n$ 
classical bits. This is often stated as a quantum analog of the 
classical one-time pad. However, the no-hiding theorem reveals an important 
difference between the classical one-time pad and the quantum analog. 
In the classical one-time pad 
when information is encrypted using a random key, then the original information 
is neither in the message nor in the key but hidden in the correlations.  
In the quantum case, when the original information is missing it simply resides 
in the key (in the unitary version in the two qubit ancilla) from which one can 
recover the original qubit. In this sense our experiment also demonstrates
an important difference between the classical one-time pad and its 
quantum analog. The no-hiding theorem not only holds in the perfect 
case (where information completely disappears) but also it is 
robust to imperfections.
%We hope that NMR setup continues to be a test bed for
%verifying quantum mechanical ideas in future.
We hope that NMR will continue to be an excellent test bed for
verification of quantum mechanical predictions and our work will 
inspire others to test the no-hiding theorem in the context of
quantum teleportation and thermalization.

\vskip 1cm

Acknowledgements: We thank P. Rungta for useful discussions.

% \end{multicols}
\end{document}